\documentclass[]{aa}
\usepackage[dvips]{graphicx}
\usepackage{natbib}
\bibpunct{(}{)}{;}{a}{}{,}
\usepackage{rotating}

\newcommand{\Msun}{M$_\odot$}
\newcommand{\Rsun}{R_\odot}
\newcommand{\Lsun}{L_\odot}
\newcommand{\mum}{12~$\mu$m}
\newcommand{\Teff}{$T_{\rm eff}$}
\newcommand{\Mbol}{$M_{\rm bol}$}

\begin{document}


\thesaurus{05     
         (10.07.2; 
          10.07.3: 47 Tuc;
          08.16.4; 08.03.4; 08.13.2; 13.09.6 )
}

\title{
Evolution of the dust mass loss with luminosity 
along the giant branch of the globular
  cluster 47 Tuc}
\author{
A. Ramdani\inst{1,2}
\and
A. Jorissen\inst{2}
\thanks{Research Associate,
F.N.R.S., Belgium}\\
}
\offprints{A. Jorissen email: ajorisse@astro.ulb.ac.be}
\institute{
D\'epartement de Physique, Universit\'e d'Oujda, Maroc
\and
Institut d'Astronomie et d'Astrophysique, Universit\'e Libre de
Bruxelles, CP 226, Boulevard du Triomphe, B-1050 Bruxelles,
Belgium
}

\date{Received / Accepted }

\titlerunning{Mass loss of giant stars in 47 Tuc}
\maketitle

\begin {abstract}
The present paper investigates the properties of the dust mass loss
in stars populating the giant branch of the globular cluster 47 Tuc, by
combining ISOCAM
and DENIS data. Raster maps of 5 fields covering areas ranging from 
$4\times4$ to $15\times15$\,arcmin$^2$ 
at different distances from the center of the cluster
have been obtained with ISOCAM at 11.5\,$\mu$m (LW10 filter). The 
covered 
fields  
include most of the red variables known in this cluster.    
A detection threshold of about 0.2 mJy is achieved, allowing to detect 
giant stars at 11.5\,$\mu$m all the way down to the horizontal branch.
No dust-enshrouded asymptotic giant branch stars 
have been found in the observed fields, contrary to the situation
encountered in LMC/SMC globular clusters with larger turnoff masses. \\
The color index $[12]-[2]$ (based on the ISO 11.5~$\mu$m flux and on the 
DENIS $K_s$ magnitude) is used as a diagnostic 
of dust emission (and hence dust mass loss). 
Its evolution with luminosity along the giant branch reveals 
that dust mass loss is only present in V3
(the only cluster Mira variable observed in the present study)
and in V18, a star presenting intermittent variability. 
This conclusion confirms
the importance of stellar pulsations in the dust formation and
ensuing mass loss.
\end{abstract}

\section{Introduction}
\label{Sect:intro}

Stellar mass loss is an essential ingredient in the modelling of several
important astrophysical processes. 
It plays a key role in the chemical evolution
of the Galaxy, by returning to the interstellar medium the ashes of the 
nuclear-burning processes that took place during the star's lifetime. 
In Asymptotic
Giant Branch (AGB) stars, the mass loss rate is so large 
that the evolutionary timescale is controlled by mass loss rather than by
nuclear burning. The general principles governing mass loss on the AGB are 
understood. Shock waves associated with the stellar pulsation push the
gas far enough above the photosphere for dust to start condensing. Radiation
pressure on the dust, when it couples to the gas, then provides the necessary
thrust for matter to escape (e.g., Sedlmayr \& 
Winters 1997).   
A parametrization of the mass loss rate on the AGB, in
terms of the fundamental stellar parameters $M, L, R$ and $Z$, is not 
available from empirical data (e.g., Zijlstra 1995), but Arndt et
al. (1997) have proposed one based on computed models (see below).  
The oldest among empirical mass-loss-rate 
parametrizations is the Reimers formula (Reimers 
1975; see Zijlstra 1995 and Sect. 11 of Habing 1996 for a discussion
of the more recent parametrizations):
\begin{equation}
\dot{M} ({\rm M}_\odot \; {\rm y}^{-1}) 
= -4\; 10^{-13}\; \eta\; \frac{(L/\Lsun) (R/\Rsun)} {(M/{\rm M}_\odot)}, 
\end{equation}
where $\eta$ is a numerical factor [introduced later
by Fusi-Pecci \& Renzini (1976) and Mengel (1976)] 
of the order of unity.
Although the Reimers formula
provides a fair
representation of the mass loss rates measured for
G, K and (non-long-period-variable) M giants,
it is well known that it predicts 
rates that are too low at the AGB-tip (e.g., de Jong 1983; 
Willson 1987; Bowen \&
Willson 1991; Bl\"ocker 1995). The final AGB masses
obtained with a  mass loss rate following the Reimers prescription 
all the way to the AGB-tip
are too large to satisfy the initial-final
mass relationship of Weidemann (1987). Therefore, Renzini (1981)
suggested the existence of a short episode of strong mass loss close to 
the AGB-tip (dubbed `superwind'), 
with mass loss rates of the order of $10^{-5}$ to
$10^{-4}$~\Msun~y$^{-1}$, whereas the Reimers formula predicts maximum 
mass loss rates of the order of $10^{-6}$~\Msun~y$^{-1}$. 
Luminous AGB stars with mass loss
rates that large have since been found (e.g., de Jong 1983; 
Whitelock et al. 1991,
1994; Vassiliadis \& Wood 1993 and references therein), as well as
empirical evidences
for a sudden increase of the mass loss rate at the end of the AGB
(e.g., Delfosse et al. 1997 and references therein). 
There has been much speculation concerning the nature and cause
of the superwind. 
Willson (1981; her Figs.~5 and 6) and Bowen \& Willson (1991) 
argued that pulsation-driven mass loss in Mira variables accounts 
for the required superwind. On the other hand, 
Netzer \& Elitzur (1993) argued that the mass loss rate abruptly
increases when the dust couples to the gas, thus causing the
superwind phase. Detailed calculations of the AGB evolution, including
all the ingredients (stellar structure, pulsation and shock waves, dust
formation and
coupling to the gas, radiative transfer) 
necessary to trigger mass loss in a self-consistent way 
are now being performed for carbon-rich Miras (Fleischer et al. 1992; H\"ofner
et al. 1996; 
Sedlmayr \& Winters 1997; Winters et al. 2000) and reproduce the
expected features of the superwind phase at the AGB-tip (Schr\"oder et 
al. 1999). A parametrization of the mass loss rates, in terms of
$T_{\rm eff}$, $L$ and $M$, and based on the
computed models, has been proposed  by Arndt et al. (1997).

The sensitivity of mass loss to
metallicity is one of the remaining question in the field. The
low-metallicity behavior of dust- and pulsation-driven AGB mass loss
has been investigated  
by Bowen \& Willson (1991) with the conclusion that low-metallicity
AGB stars reach the same mass loss rates than their higher
metallicity counterparts, albeit at higher luminosities.
This prediction  is basically confirmed by 
observations of mass-losing AGB stars in the
Magellanic Clouds, which  do not call for low mass loss rates at
low $Z$ (see Zijlstra 1999 and Olofsson 1999 for recent reviews).    

Giant stars in globular clusters are ideal targets to address  
these questions.
In particular, the variation of mass loss
with luminosity may be easily followed, the impact on mass loss of the 
pulsation properties of the long-period
variables  may be investigated,  as well as the impact of the 
subsolar metallicity characterizing globular-cluster stars.

Among globular clusters, 47 Tuc (NGC 104) offers several advantages. It hosts
many well-studied long-period variables (Sawyer-Hogg 1973, Glass \& 
Feast 1973, Lloyd Evans 1974, Fox 1982, Frogel 1983), among which 4
Mira variables (V1 to V4) with periods ranging from 165 to 212~d, and
visual amplitudes ranging from 1.5 to 4 mag. 
It is nearby (distance modulus = 13.7; Gratton et al. 1997)
and has been the subject of many previous studies providing auxiliary data like
bolometric magnitudes from $UBVIJHKL$ fluxes or assignments of its giant stars
to either the AGB 
or the RGB (Frogel et al. 1981). The basic parameters of 47~Tuc are as
follows: [Fe/H]$ = -0.67$, age $\sim 10.5$~Gy   
(Gratton et al. 1997), turnoff mass $\sim 0.9$~\Msun\ (Hesser et al. 1987).
Gillett et al. (1988) have performed a pointed observation of 47 Tuc with 
IRAS, and detected several sources at 12, 25 and 60 $\mu$m (see
Table~\ref{Tab:IRASISO}). 
The use of ISOCAM onboard ISO (see Sect.~\ref{Sect:observations})
provides both a much better sensitivity (0.2~mJy instead of 10~mJy at
\mum) and a better angular
resolution (a few arcseconds instead of 45 arcsec) than IRAS, and thus
makes it possible to detect many more faint point sources than it was
possible with IRAS.

The present study aims at monitoring the dust mass loss as a function
of luminosity 
along the RGB and AGB in 47 Tuc. Relatively warm dust is
easily detected from the flux excess that it produces at 12
$\mu$m. Several authors (e.g., Whitelock et al. 1994, Le Bertre \& Winters
1998, Jorissen \& Knapp 1998) have shown that there is a tight correlation
 (at least for oxygen-rich stars in the solar neighbourhood) between the mass
loss rate and the
$K - [12]$ color index, which may thus be used to trace the dust mass 
loss. 

Another objective of the present study is to detect possible  dust-enshrouded
stars with no optical counterparts, as found by Tanab\'e et al. (1997, 1999) in
globular clusters from the Magellanic Clouds. 

\section{Observations}
\label{Sect:observations}

Mosaic images of 5 fields at different distances 
from the center of the 47 Tuc cluster  have been obtained with the
ISOCAM camera on board ISO 
\citep[see][for descriptions of ISO and ISOCAM
respectively]{Kessler96,Cesarsky96}, 
using the LW10 
filter (closely matching the IRAS 12\,$\mu$m filter). The 
field positions have been chosen so as to survey most of the red 
variable stars
in 47
Tuc. The 
Astronomical Observation Template (AOT) parameters for each of the fields are
given in Table~\ref{Tab:AOT}. They were adjusted so as 
to detect sources as faint as possible without saturating on the
brightest sources. 
These contradictory requirements made us select a number $N_{\rm obs}$ 
of elementary integrations as large as possible (namely $N_{\rm obs} = 
10$, to keep within the total allocated time) but with short integration times 
($T_{\rm int} = 2$~sec) and low gain ($\times1$). With this setup, the
detector saturates for fluxes of the order of 0.8~Jy/pixel, and remains
linear up to about 0.3~Jy/pixel.
The pixel field of view was taken equal to either $3''$ (with
the large
field mirror) for the fields closest to the core, or $6''$ for fields farther
out. The step size between two successive pointings of the mosaic was half 
the chip size (in either direction), i.e., $90''$ for the $6''$
pixel field of view or $45''$ for the $3''$
pixel field of view, to ensure 4 independent detections of the
sources in the central region of the mosaic. The total integration time per
source is therefore $4 \times N_{\rm obs} \times T_{\rm int}$. The very center
of the cluster has not been observed to avoid saturation and crowding problems.

\begin{sidewaystable*}
\caption[]{\label{Tab:AOT}
Astronomical Observation Template parameters for the
  47 Tuc observations (ISO program `AJORISSE 47TUC') }

\begin{tabular}[]{lllllllllllllllllllll}
\hline\\
Filter & \multicolumn{5}{c}{LW-10 (center 11.5\,$\mu$m, 
range 8.0-15.0\,$\mu$m)}
\cr
Date of observation & \multicolumn{5}{c}{25-May-1996}\cr
\hline\\
\hline\\
Field name &          FIELD2 & V5 &  FIELD3 & FIELD4 & FIELD1 \\ 
\hline
Planned targets &V3,V7,V11,V15,V18  & V5 & Lee \#1421 & V13
&Lee \# 5529 \\ 
Field center (J2000): $\alpha$ & 0h25m35s  &  0h25m08s & 
0h22m46s & 0h22m47s & 0h27m0s\\ 
\hspace*{3cm} $\delta$   & $-72^\circ04'30''$  &
$-72^\circ10'15''$ & $-72^\circ18'00''$ & $-72^\circ09'07''$ &
$-71^\circ55'00''$ \\  
Distance from cluster center (arcmin) & 6.9  & 7.2 & 14.5 & 7.4 
& 16.7 \cr 
Detector gain            & 1              & 1 & 1 & 1 & 2\cr 
Pixel Field of View (arcsec) & 3          & 3 & 6 & 3 & 6\cr 
Number of pointings & 9$\times$9          & 4$\times$4 &
9$\times$9 & 7$\times$7 & 8$\times$8\cr 
Total field of view (arcmin$^2$) & 7.6 $\times$ 7.6            
& 3.85 $\times$ 3.85 & 15.2 $\times$ 15.2 & 6.1 $\times$ 6.1 
& 13.7 $\times$ 13.7\cr 
Step between two pointings (arcsec)& 45    & 45 & 90 & 45 & 90\cr 
\hline 
\end{tabular}
\end{sidewaystable*}

\begin{sidewaystable*}[ht]
\caption[]{\label{Tab:IRASISO}
Comparison of IRAS and ISOCAM 12\,$\mu$m fluxes.
IRAS fluxes are from Gillett et al. (1988), and $K$ magnitudes from Frogel et
 al. (1981), Glass \& Feast (1973) or DENIS (Cioni et al. 2000). The
variable name refers to Sawyer-Hogg (1973), and the four-digits number
for non-variable stars refers to Lee (1977). The period $P$ and visual
amplitude $\Delta V$ for the variable stars are taken from Sawyer-Hogg
(1973) or Fox (1982)}
\renewcommand{\arraystretch}{0.7}
\begin{tabular}{ccccccccclcll}
\hline\\[-5pt]
Name   & Alternate    & \multicolumn{2}{c}{$F(12)$ (Jy)} &&
\multicolumn{3}{c}{$K$} &$F12/F2$ & $P$ & $\Delta V$ & Rem.\cr
\cline{3-4}\cline{6-8}
     & name & IRAS & ISOCAM && Frogel/Glass & DENIS & DENIS name & &
     (d)  & (mag) \\[+5pt]
\hline\\[-5pt]

V3  &      & 0.215  & 0.203 && 6.00 -- 6.49 & --  & & 0.08 -- 0.12 &
192 & 4.15 \cr
V5  &      & 0.028  & 0.038 && 7.47 & 7.35 & J002501.97-720929.9 &
0.056 & 45, 60--70 & 0.7 \cr 
V7  &      &  --    & 0.055 && 6.97 & 6.79 & J002528.42-720652.2 &
0.051 & 50--58 & 0.7 \cr
V13 &      & 0.034  & 0.049:&& 7.65 &      &  & - & 40 & 0.7 &
\multicolumn{1}{c}{*}\cr
V15 & W300 & 0.041  & 0.043 && 7.27 & 7.03 & J002551.35-720703.8 &
0.052 & 38 & 0.2 \cr
V11 & W12  &   --   & 0.095 && 6.69 & 6.48 & J002506.64-720220.7 &
0.068 & 52:, 100 & 0.8 \cr
V18 & L168 &   --   & 0.105 && 7.45 & --   & & 0.150 & & 0.3 \cr
V11+V18 &  & 0.235  & 0.200 &&  --  &  --  &  --    
\bigskip\cr  
7701& R17  &   --   & 0.028 && 7.66 & 7.43 & J002532.43-721052.2 & 0.049 \cr 
8745& R18=V16&   --   & 0.036 && 7.44 & 7.08 & J002529.74-721118.4 & 0.051  \cr
7701+8745& & 0.065  & 0.064 &&  --  &  -- 
\bigskip\cr
R19 &      & 0.034  & 0.032:&&      &      &      & & &  &
\multicolumn{1}{c}{*}\cr
V28 & LR5  & 0.078  & 0.071 &&      &      &      & & 40: & 0.4 & FIELD3, *\cr
    &      &        & 0.065 &&      &      &      & & &     & FIELD4\cr
2620&      & 0.018  & 0.021 &&      &      &     \cr
5529&      & 0.020  & 0.022 && 7.92 & 7.78 & J002604.55-715324.6 & 0.048 \cr
6304&      & 0.031  & 0.032 &&      & 7.22 & J002738.23-715257.6 & 0.038 \cr
1421&      & 0.071  & 0.062 && 6.84 & 6.68 & J002330.55-722236.5 & 0.050 \cr
1601&      & 0.023  & 0.024 &&      & 7.69 & J002401.03-721513.0 & 0.044 \cr
1603&      & 0.026  & 0.018 && 7.94 & 7.85 & J002339.15-721639.9 & 0.042  \cr
1533&      & 0.026  & 0.013 &&      & 8.21 & J002357.90-721857.8 & 0.038  \cr
\hline 
\end{tabular}


Remarks:\\
V13: The ISOCAM image is a blend of 3 sources, as seen from the 
{\it Digital Sky Survey} image\\
R19: The ISOCAM image is a blend of 2 sources, as seen from the 
{\it Digital Sky Survey} image\\
1421: Lee (1977) suspects this star to be a LPV\\
V 28 = LR 5 = Lee \#2758
\end{sidewaystable*}

\section{Reductions}
\subsection{ISOCAM image}
\label{Sect:ISOCAM}

The reduction of our ISOCAM data poses a special challenge as we are aiming at
extracting sources as faint as possible in a field containing very bright (AGB)
sources. Because of the strong detector memory effects, these sources will    
leave ghost images in the next pointings that ought not to be confused with 
faint sources. The PRETI routines allowed us to overcome at least some 
of these
problems. They were designed by Starck et al. (1999; see also Aussel et al.
1999) specifically for the detection of faint sources on ISOCAM images, after
performing a careful removal of the cosmic ray glitches (both their short- and
long-term effects) and correcting for the transient behaviour so as to provide
a flat baseline.   

The main steps of the reduction process are as follows.
The default dark frame is first subtracted from the raw data 
(CIER files) using the CAM Interactive Analysis (CIA) package 
(Ott et al. 1997).
The bad column 23 is replaced by the average of columns 22 and 24. 
The short-duration cosmic rays are identified from a 3$\sigma$ test on the
short-term
fluctuations of successive readouts and the affected readouts are masked. 
The long-term effects of cosmic rays are identified from their
typical pattern in a multi-resolution median filter, and masked as well.
Remaining long-term drifts (mainly due to the detector memory effect after
passing on bright sources, and to the change of the ISOCAM configuration at the
beginning of the observation) are suppressed by subtracting the baseline of 
each
pixel's time history, obtained from the sequence of all the individual images. 
  
The flat field is constructed from the median of the data
themselves, since in the best cases, about 800 ($\sim 10 \times 9 \times 9$) 
readouts are available per pixel.
The detector transient is corrected by the `inverse method' developed by 
Abergel et al. (1996).
In the present analysis, the fluctuations arising on top of the bright sources
(probably caused by the jitter of the satellite) are often confused with short-
term cosmic rays and unduly masked, resulting in a loss of useful readings
for bright sources. 
This problem could possibly lead to underestimating the flux of bright sources,
but the effect - if present - does not appear to be very severe, as will be seen
from a comparison of our ISOCAM fluxes with previous IRAS measurements of the
same sources (see Table~\ref{Tab:IRASISO}). 

\begin{figure*}[!ht]
\raisebox{-8cm}{
\begin{minipage}[]{9.0cm}
\begin{picture}(9.0,10.8)
\resizebox{\hsize}{!}{\includegraphics{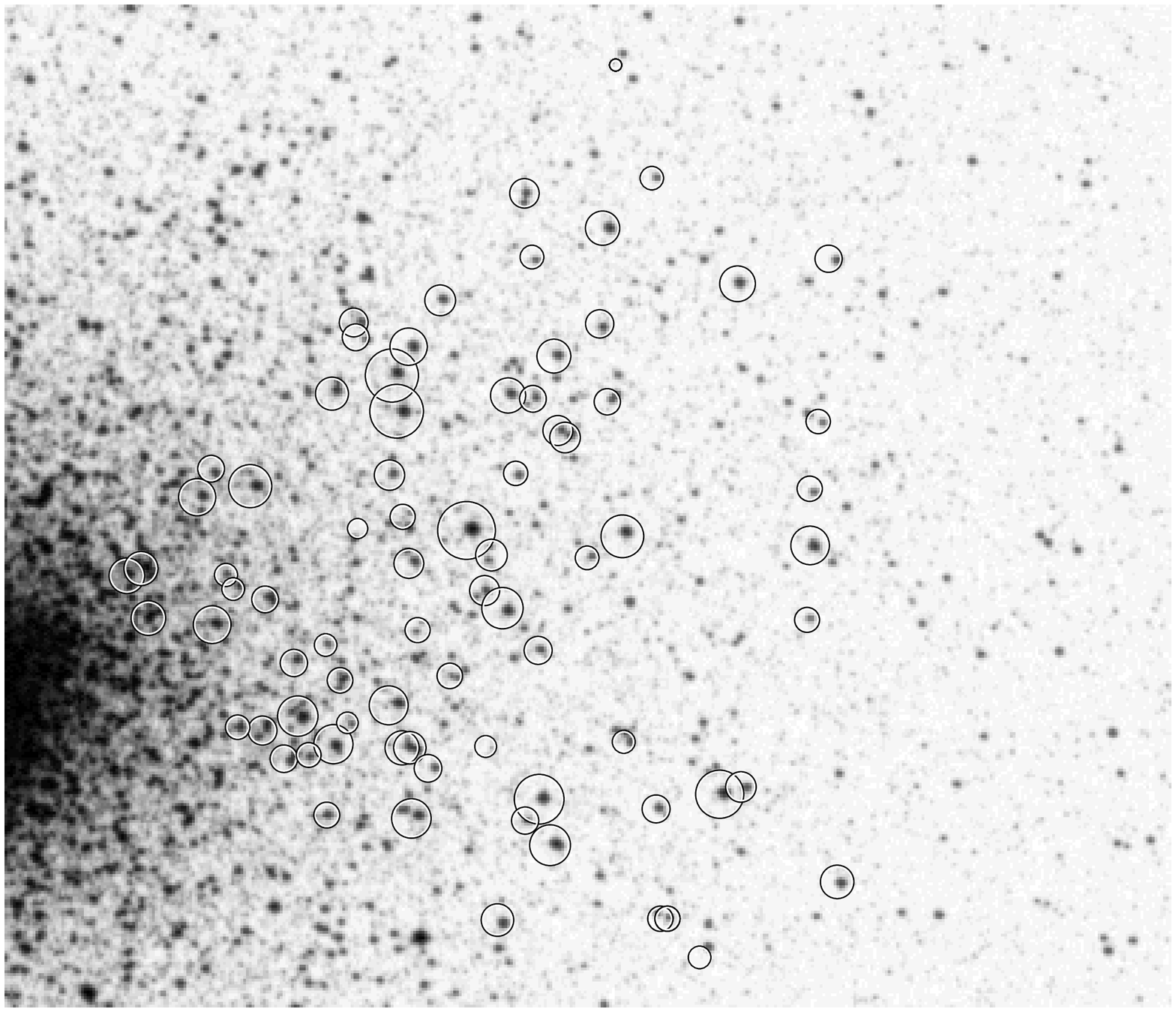}}
\end{picture}
\end{minipage}
}
\hfill
\raisebox{-4cm}{
\begin{minipage}[]{8.5cm}
\begin{picture}(8.5,10)
\resizebox{\hsize}{!}{\rotatebox{135}{\includegraphics{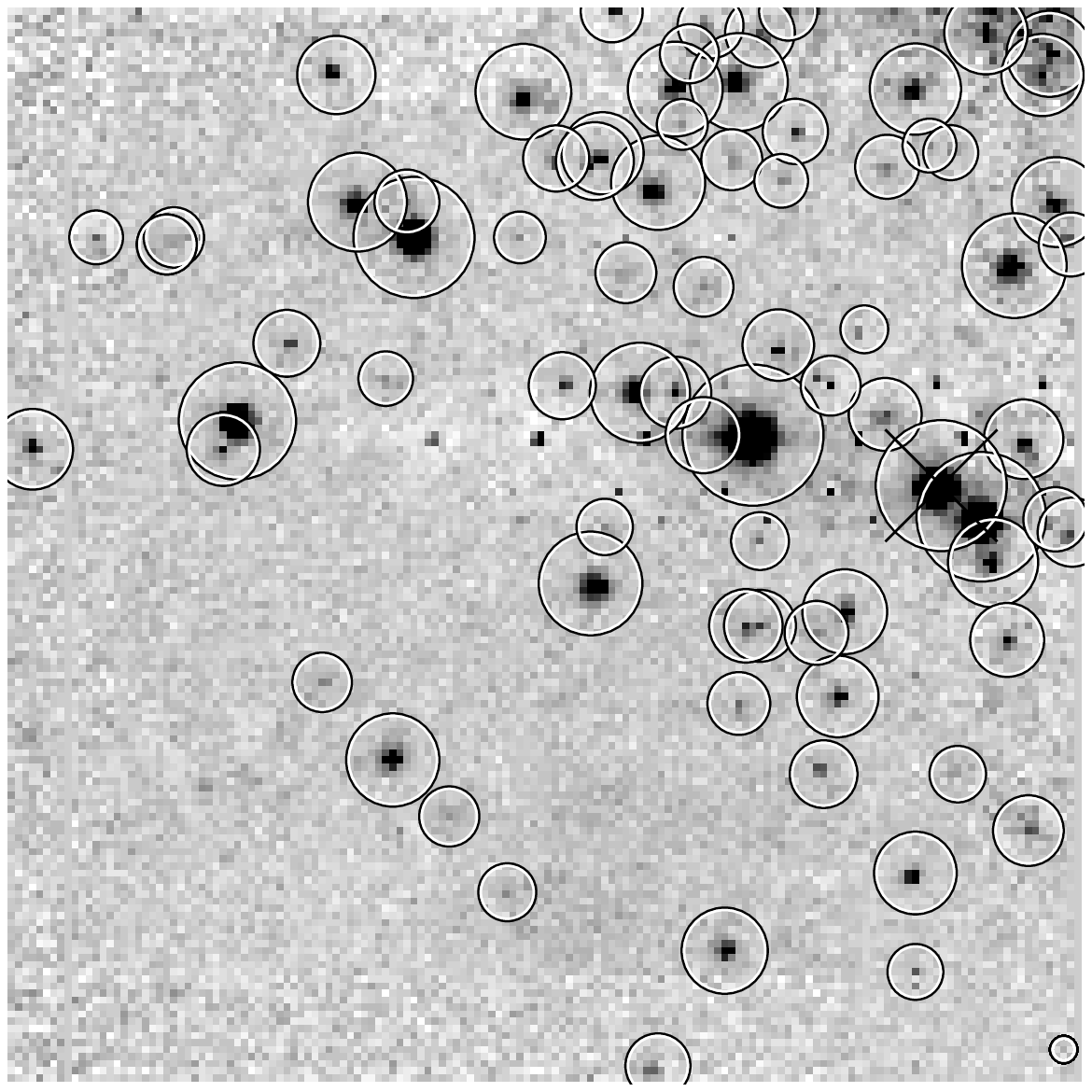}}}
\end{picture}
\end{minipage}
}
\caption{\label{Fig:image}
Comparison of the {\it Digital Sky Survey} (left panel) and ISOCAM
12~$\mu$m 
(right panel) images 
corresponding to FIELD2. 
The 12 $\mu$m sources
are identified by circles with a radius proportional to their 12 $\mu$m
flux (the variable V18
discussed in the text is identified by the cross on the right panel). 
East is to the right, North is up. The field size of the {\it
  Digital Sky Survey} image is 12 $\times$ 12 arcmin$^2$
}
\end{figure*}

The individual mosaic frames are then coadded into a single image. 
The source detection is performed on that 
image (Fig.~\ref{Fig:image}) using a wavelet transform, which detects at each
scale all pixels
of the image above a given threshold $\sigma$ of the noise map (Starck et al.
1999).
A good compromise between maximum sensitivity  and minimum number of false
detections (as compared to optical sources; see
Sect.~\ref{Sect:optical}) 
is obtained with
a detection threshold set at 5$\sigma$. More details about the
criteria of inclusion of sources in the
final source list are provided in Sect.~\ref{Sect:list}.

The source fluxes are obtained by aperture photometry on the reduced
image. The aperture used includes 12 pixels, and is the sum of the 
3 $\times$ 3-pixels square centered on
the source, plus 4 pixels exterior to this square and located 
in the middle of each side of the square. 

The measured values were then multiplied by the factors 100/72 (for
the fields observed with a $3''$ pixel field-of-view; see
Table~\ref{Tab:AOT}) or
100/84.5 (for
the fields observed with the $6''$ pixel field-of-view) to account for
the point-spread-function (PSF) 
 wing not falling on the selected aperture. Those factors were 
computed from the observed PSF provided by Okumura 
(ISOCAM PSF Report, 04/Nov/1998, available on the ISOCAM page\\ 
http://www.iso.vilspa.esa.es/users/expl\_lib/CAM\_list.html).

The measured ADUs are converted to Jansky units using the 
conversion factor provided by the  CIA software package
(4.129~ADU/sec/mJy/pix).
The uncertainty on the \mum\ flux is the quadratic average of the Poisson
noise on the signal, 
and of the uncertainty associated with
the PSF correction (3\%).

Because the reduction process is an intricated one implying many different
steps, it is interesting to compare the ISOCAM fluxes with previous results
obtained by Gillett et al. (1988) from pointed IRAS observations of 47~Tuc
(Table~\ref{Tab:IRASISO}).
 
The 12~$\mu$m IRAS and ISOCAM fluxes generally agree within 25\%, which is
satisfactory considering the fact that the filters are not identical and that
some of the LPV sources may be variable at 12~$\mu$m as well.
The largest deviations  are obtained for the composite source V11+V18 (18\%),
and for V13 (31\%). The \mum\ flux of V13 is not reliable, as the
ISOCAM image of V13 is in fact a blend of 3 different stars
that are seen separately on the {\it Digital Sky Survey} image
(Sect.~\ref{Sect:optical}). Another discrepancy concerns V7, which
does not seem to have been detected by Gillett et al. (1988). 
This non-detection
is surprising, since its flux
at the time of our ISOCAM observation is similar to that of V15 which was
detected by
Gillett et al. (1988). A possible explanation would be that V7 was much fainter
in the
\mum\ band at the time of the IRAS observation. This seems unlikely, however,
since the  
ISOCAM flux for V7 agrees with the prediction for a dust-free photosphere (see
Fig.~\ref{Fig:F12F2}), so that a smaller IRAS flux would mean that V7 was then
underluminous at \mum.

Note that the star V28 (= LR 5 = Lee \#2758) 
is present on both FIELD3 and FIELD4. The
ISOCAM fluxes from these two independent measurements differ by only
9\% (Table~\ref{Tab:IRASISO}), 
which provides an estimate of the internal consistency of the 
derived fluxes.

\subsection{DENIS data}
\label{Sect:DENIS}

As the $K_s$ magnitude provided by the {\it Deep Near Infrared Survey} (DENIS;
see Epchtein 1998) saturates at $K_s = 6.5$ (Epchtein 1998),
$K$ magnitudes for the brightest sources detected on the ISOCAM images
are taken from Frogel et al. (1981) or from Glass \& Feast (1973).
For the fainter sources, data from DENIS have been used, as published by Cioni
et al. (2000) for
fields around the Small Magellanic Cloud. 

The cross-identification between ISOCAM and DENIS sources has been performed by
overplotting their positions on the ISOCAM image. In most cases, an almost 
constant offset separated the positions of a given source in the two datasets. 
In some cases,
however, an atypical offset whose origin is currently unknown 
had to be applied, especially for the
brightest 
sources (V7 and V15). The cross-identification of these sources
makes little doubt, as they generally correspond to the brightest sources
in the area. Examples of cross
identifications are listed in Table~\ref{Tab:IRASISO}.

The variable V3 is the only one to vary by a large factor (about 4 mag 
in $V$ and 0.5
mag in  $K$; Sawyer-Hogg 1973, Frogel et 
al. 1981). The other variables have amplitudes smaller than  
0.9 mag in $V$ (Lloyd Evans 1974, Fox 1982). Therefore, they are not
expected to vary by more than 0.1  mag in $K$. 
The average deviation $K({\rm Frogel}) - K_s({\rm DENIS})$
is 0.21 mag (excluding V3).

\subsection {Optical data}
\label{Sect:optical}

\begin{figure}
\resizebox{\hsize}{!}{\includegraphics{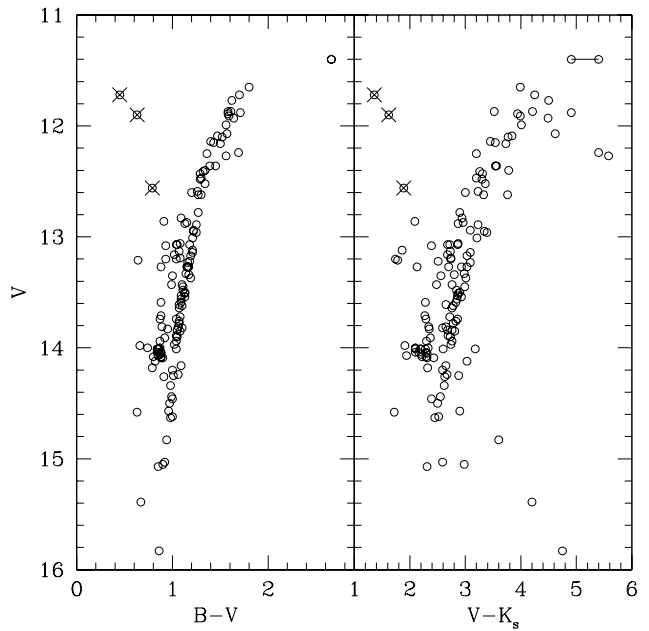}}
\caption[]{
The $(V, B-V)$ (left panel) and $(V, V-K_s)$ (right panel) 
diagrams for the stars detected at 12~$\mu$m and with $B$,
$V$ magnitudes available from Lee (1977) or Chun \& Freeman
(1978), and $K_s$ from Table~\protect\ref{Tab:sources}.
The outliers (non-members?) 1321, 1412, 6209 
are denoted by crosses 
}
\label{Fig:VBV}
\end{figure}

Lee (1977) and Chun \& Freeman (1978) 
provide $UBV$ data for most of the fields surveyed by the
present ISOCAM observations. The assignment of a given giant star to either 
the RGB or AGB is possible from the   
$(V, B-V)$ diagram presented in Fig.~\ref{Fig:VBV} for the 
stars detected at 12~$\mu$m. 

Optical images from the {\it Digital Sky Survey} were used to identify
the optical counterparts of the  infrared sources, along with the optical
positions provided by Tucholke (1992) and its cross-identifications with the
catalogues of Lee (1977) and Chun \& Freeman (1978).
The left panel of Fig.~\ref{Fig:image} 
identifies the 12 $\mu$m sources on the optical  image for FIELD2. 
Basically all sources brighter than $V = 14$ (i.e.\ $M_V < 0.3$ for 
the distance modulus $V - M_V = 13.7$ of 47 Tuc) 
are detected at 12 $\mu$m, corresponding to a completeness threshold
of about 0.7 mJy including
all the giants down to the horizontal branch (Fig.~\ref{Fig:VBV}).

\section{Final ISOCAM source list and absence of dust-enshrouded AGB stars}
\label{Sect:list}

\begin{table*}
\caption[]{
\label{Tab:sources}
List of sources detected at 12 $\mu$m. The first column provides the source
identifier from Lee (1977), Chun \& Freeman (1978) or Tucholke (1992) (the
latter being referred to by `tuc' followed by a 4-digit number). The next three
columns list the $V$ magnitude
from Lee (1977) or Chun \& Freeman (1978), the DENIS $K_s$ magnitude (except 
for V3, where $K$ is taken from Frogel et al. 1981), and the
$V-K$ index. The fifth column lists the magnitude at 12 $\mu$m, derived from
the flux in Jy ($F12$) using the relation $[12] = -2.5 \log F12/28.3$.
The $F12/F2$ ratio is listed next, with $F2({\rm Jy}) = 665\; 10^{-0.4
  K_s}$. 
The uncertainty $\sigma(F12/F2)$ on $F12/F2$ is given in
column~7, and includes only the uncertainty coming from $F12$ (see
Sect.~\ref{Sect:ISOCAM}). Column~8 indicates the field number to which
the star belongs, and the last column contains remarks (like
non-membership, based on an outlying position in the $(V, B-V)$ 
or $(K_s, V-K_s)$ diagrams)
}
\renewcommand{\arraystretch}{0.9}
\begin{tabular}{rrrrrrrrr}\hline
Star &$V$ &$K_s$&$V-K$&[12]&  $F12/F2$ &$\sigma(F12/F2)$ & Field Number \\ \hline
6304   & 8.51   &   7.22   &   1.29   &   7.35   &   0.038   &   0.003  & F1 & non member\\ 
1601   & 8.88   &   7.69   &   1.19   &   7.67   &   0.044   &   0.005  & F3 & non member\\ 
1533   & 10.09   &   8.21   &   1.88   &   8.36   &   0.038   &   0.005  & F3 &non member\\ 
d-502,V3   & 11.40   &   6.49   &   4.90   &   5.36   &   0.123   &   0.006  & F2 \\ 
d-502,V3   & 11.40   &   6.00   &   5.40   &   5.36   &   0.076   &   0.003  & F2 \\ 
d-230,V13   & 11.65   &   7.66   &   3.99   &   6.90   &   0.085   &   0.007  & F4 \\ 
1412   & 11.72   &   10.36   &   1.36   &   10.64   &   0.033   &   0.013  & F3& non member \\ 
8756,V5   & 11.72   &   7.47   &   4.25   &   7.17   &   0.056   &   0.005  & F5 \\ 
7726,V15   & 11.77   &   7.27   &   4.50   &   7.05   &   0.052   &   0.004  & F2 \\ 
d-596   & 11.87   &   8.35   &   3.52   &   8.35   &   0.043   &   0.006  & F2 \\ 
2620   & 11.87   &   -   &   -   &   7.84   &   -   &   -  & F4 \\ 
7701,R17   & 11.87   &   7.66   &   4.21   &   7.50   &   0.049   &   0.005  & F5 \\ 
e-484   & 11.88   &   -   &   -   &   7.94   &   -   &   -  & F2 \\ 
d-598,V7   & 11.88   &   6.97   &   4.91   &   6.78   &   0.051   &   0.004  & F2 \\ 
1603   & 11.89   &   7.95   &   3.94   &   7.96   &   0.042   &   0.005  & F3 \\ 
1321   & 11.90   &   10.28   &   1.62   &   10.18   &   0.048   &   0.015  & F3& non member \\ 
5529   & 11.91   &   7.92   &   3.99   &   7.78   &   0.048   &   0.005  & F1 \\ 
8745,R18   & 11.93   &   7.44   &   4.49   &   7.24   &   0.051   &   0.005  & F5 & \\ 
2758,LR5   & 11.94   &   -   &   -   &   6.50   &   -   &   -  & F3 \\ 
2758,LR5   & 11.94   &   -   &   -   &   6.59   &   -   &   -  & F4 \\ 
6747   & 11.99   &   7.98   &   4.01   &   8.01   &   0.042   &   0.005  & F2 \\ 
2721   & 12.06   &   -   &   -   &   8.43   &   -   &   -  & F4 \\ 
d-493,V18   & 12.07   &   7.45   &   4.62   &   6.08   &   0.150   &   0.009  & F2 \\ 
e-558   & 12.08   &   -   &   -   &   8.42   &   -   &   -  & F2 \\ 
d-509   & 12.09   &   8.25   &   3.84   &   8.26   &   0.043   &   0.006  & F2 \\ 
2426   & 12.10   &   8.33   &   3.77   &   8.25   &   0.046   &   0.006  & F3 \\ 
e-563   & 12.14   &   8.69   &   3.45   &   8.64   &   0.046   &   0.007  & F2 \\ 
1510   & 12.15   &   8.61   &   3.54   &   8.81   &   0.036   &   0.006  & F3 \\ 
1505   & 12.16   &   8.43   &   3.73   &   8.55   &   0.038   &   0.006  & F3 \\ 
5312   & 12.16   &   -   &   -   &   8.36   &   -   &   -  & F1 \\ 
2734   & 12.21   &   -   &   -   &   8.68   &   -   &   -  & F4 \\ 
e-599   & 12.23   &   -   &   -   &   8.58   &   -   &   -  & F2 \\ 
1421   & 12.24   &   6.84   &   5.40   &   6.65   &   0.050   &   0.003  & F3 \\ 
6719   & 12.24   &   -   &   -   &   9.50   &   -   &   -  & F2 \\ 
\hline
\end{tabular}

\end{table*}

\addtocounter{table}{-1}
\begin{table*}
\caption[]{(Continued).\\
}
\begin{tabular}{rrrrrrrrr}\hline
Star &$V$ &$K_s$&$V-K$&[12]&  $F12/F2$ &$\sigma(F12/F2)$ & Field Number \\ \hline
d-141   & 12.25   &   9.05   &   3.20   &   9.18   &   0.038   &   0.008  & F4 \\ 
d-492,V11   & 12.27   &   6.69   &   5.58   &   6.18   &   0.068   &   0.004  & F2 \\ 
6504   & 12.28   &   -   &   -   &   10.86   &   -   &   -  & F1 \\ 
6764   & 12.30   &   -   &   -   &   8.75   &   -   &   -  & F2 \\ 
1709   & 12.36   &   8.80   &   3.56   &   8.89   &   0.040   &   0.007  & F4 \\ 
1709   & -   &   -   &   -   &   9.24   &   0.029   &   -  & F3 \\ 
1605   & 12.36   &   8.82   &   3.54   &   8.71   &   0.047   &   0.008  & F3 \\ 
6527   & 12.39   &   -   &   -   &   9.02   &   -   &   -  & F1 \\ 
e-538   & 12.40   &   8.62   &   3.78   &   8.68   &   0.040   &   0.007  & F2 \\ 
1513   & 12.41   &   9.15   &   3.26   &   9.05   &   0.047   &   0.009  & F3 \\ 
2525   & 12.43   &   9.12   &   3.31   &   9.22   &   0.039   &   0.008  & F3 \\ 
5422   & 12.44   &   -   &   -   &   8.97   &   -   &   -  & F1 \\ 
6768   & 12.47   &   9.27   &   3.20   &   9.25   &   0.044   &   0.009  & F2 \\ 
5627   & 12.48   &   9.18   &   3.30   &   9.73   &   0.026   &   0.007  & F1 \\ 
8740   & 12.52   &   9.16   &   3.36   &   9.43   &   0.033   &   0.008  & F5 \\ 
e-487   & 12.54   &   -   &   -   &   8.99   &   -   &   -  & F2 \\ 
6209   & 12.56   &   10.67   &   1.89   &   11.01   &   0.031   &   0.014  & F1 & non member?\\ 
d-571   & 12.59   &   9.36   &   3.23   &   9.81   &   0.028   &   0.008  & F2 \\ 
1301   & 12.60   &   9.60   &   3.00   &   9.50   &   0.047   &   0.011  & F3 \\ 
e-512   & 12.62   &   8.86   &   3.76   &   8.89   &   0.041   &   0.007  & F2 \\ 
6732   & 12.62   &   9.29   &   3.33   &   9.35   &   0.041   &   0.009  & F2 \\ 
5636   & 12.68   &   -   &   -   &   10.34   &   -   &   -  & F1 \\ 
7723   & 12.75   &   -   &   -   &   9.64   &   -   &   -  & F2 \\ 
6407   & 12.78   &   -   &   -   &   9.85   &   -   &   -  & F1 \\ 
6408   & 12.78   &   -   &   -   &   9.75   &   -   &   -  & F1 \\ 
6728   & 12.78   &   9.88   &   2.90   &   9.57   &   0.058   &   0.014  & F2 \\ 
6519   & 12.81   &   -   &   -   &   9.62   &   -   &   -  & F1 \\ 
1527   & 12.83   &   9.89   &   2.94   &   10.17   &   0.033   &   0.010  & F3 \\ 
d-143   & 12.86   &   10.77   &   2.09   &   10.99   &   0.035   &   0.016  & F4 \\ 
6717   & 12.87   &   9.91   &   2.96   &   10.12   &   0.036   &   0.011  & F2 \\ 
2742   & 12.88   &   10.01   &   2.87   &   10.25   &   0.034   &   0.011  & F4 \\ 
2742   & -   &   -   &   -   &   10.16   &   0.037   &   -  & F3 \\ 
2712   & 12.89   &   9.66   &   3.23   &   9.75   &   0.039   &   0.010  & F4 \\ 
e-561   & 12.91   &   -   &   -   &   9.53   &   -   &   -  & F2 \\ 
f-384   & 12.91   &   -   &   -   &   9.26   &   -   &   -  & F2 \\ 
5427   & 12.92   &   -   &   -   &   10.04   &   -   &   -  & F1 \\ 
1628   & 12.94   &   9.85   &   3.09   &   9.88   &   0.042   &   0.012  & F3 \\ 
e-479   & 12.95   &   9.61   &   3.34   &   9.72   &   0.039   &   0.010  & F2 \\ 
d-208   & 12.96   &   9.57   &   3.39   &   9.94   &   0.031   &   0.009  & F4 \\ 
1604   & 13.01   &   9.80   &   3.21   &   9.95   &   0.038   &   0.011  & F3 \\ 
\hline
\end{tabular}

\end{table*}

\addtocounter{table}{-1}
\begin{table*}
\caption[]{(Continued).
}
\begin{tabular}{rrrrrrrrr}\hline
Star &$V$ &$K_s$&$V-K$&[12]&  $F12/F2$ &$\sigma(F12/F2)$ & Field Number \\ \hline
1518   & 13.06   &   10.19   &   2.87   &   10.42   &   0.035   &   0.012  & F3 \\ 
1708   & 13.07   &   10.21   &   2.86   &   10.03   &   0.050   &   0.015  & F4 \\ 
1708   & -   &   -   &   -   &   10.04   &   0.050   &   -  & F3 \\ 
2604   & 13.07   &   10.39   &   2.68   &   10.45   &   0.040   &   0.014  & F3 \\ 
2741   & 13.07   &   10.35   &   2.72   &   10.56   &   0.035   &   0.013  & F4 \\ 
2741   & -   &   -   &   -   &   10.49   &   0.037   &   -  & F3 \\ 
6524   & 13.08   &   -   &   -   &   10.19   &   -   &   -  & F1 \\ 
5530   & 13.08   &   -   &   -   &   10.60   &   -   &   -  & F1 \\ 
1319   & 13.08   &   10.69   &   2.39   &   11.08   &   0.030   &   0.014  & F3 \\ 
f-394   & 13.11   &   -   &   -   &   9.44   &   -   &   -  & F2 \\ 
e-525   & 13.11   &   -   &   -   &   10.29   &   -   &   -  & F2 \\ 
7707   & 13.12   &   11.26   &   1.86   &   10.67   &   0.029   &   0.012  & F5 \\ 
1629   & 13.13   &   10.40   &   2.73   &   10.64   &   0.034   &   0.013  & F3 \\ 
d-229   & 13.14   &   10.05   &   3.09   &   9.71   &   0.058   &   0.015  & F4 \\ 
f-804   & 13.14   &   -   &   -   &   9.53   &   -   &   -  & F2 \\ 
6403   & 13.16   &   -   &   -   &   11.29   &   -   &   -  & F1 \\ 
2643   & 13.16   &   10.48   &   2.68   &   10.97   &   0.027   &   0.012  & F3 \\ 
1316   & 13.17   &   10.14   &   3.03   &   10.32   &   0.036   &   0.012  & F3 \\ 
d-144   & 13.19   &   10.45   &   2.74   &   10.59   &   0.037   &   0.014  & F4 \\ 
e-576   & 13.20   &   11.46   &   1.74   &   10.42   &   0.111   &   0.040  & F2 \\ 
1309   & 13.20   &   10.46   &   2.74   &   10.55   &   0.039   &   0.015  & F3 \\ 
6531   & 13.21   &   -   &   -   &   11.32   &   -   &   -  & F1 \\ 
1422   & 13.21   &   11.43   &   1.78   &   10.81   &   0.077   &   0.033  & F3 \\ 
2739   & 13.22   &   10.71   &   2.51   &   10.91   &   0.035   &   0.016  & F4 \\ 
2739   & -   &   -   &   -   &   10.71   &   0.042   &   -  & F3 \\ 
2428   & 13.23   &   10.14   &   3.09   &   10.32   &   0.036   &   0.012  & F3 \\ 
e-533   & 13.23   &   -   &   -   &   10.73   &   -   &   -  & F2 \\ 
d-504   & 13.23   &   -   &   -   &   10.25   &   -   &   -  & F2 \\ 
1602   & 13.24   &   -   &   -   &   10.56   &   -   &   -  & F3 \\ 
6727   & 13.27   &   10.34   &   2.93   &   10.56   &   0.035   &   0.013  & F2 \\ 
1506   & 13.27   &   10.24   &   3.03   &   10.33   &   0.039   &   0.013  & F3 \\ 
6739   & 13.27   &   10.57   &   2.70   &   10.15   &   0.062   &   0.020  & F2 \\ 
8735   & 13.27   &   11.14   &   2.13   &   11.66   &   0.027   &   0.017  & F5 \\ 
d-560   & 13.31   &   -   &   -   &   10.60   &   -   &   -  & F2 \\ 
1714   & 13.33   &   10.35   &   2.98   &   10.62   &   0.034   &   0.013  & F4 \\ 
1714   & -   &   -   &   -   &   10.62   &   0.034   &   -  & F3 \\ 
6711   & 13.34   &   10.54   &   2.80   &   10.35   &   0.051   &   0.018  & F2 \\ 
1743   & 13.35   &   10.79   &   2.56   &   10.90   &   0.039   &   0.017  & F3 \\ 
6616   & 13.36   &   -   &   -   &   10.50   &   -   &   -  & F1 \\ 
2744   & 13.37   &   -   &   -   &   11.19   &   -   &   -  & F4 \\ 
\hline
\end{tabular}

\end{table*}

\addtocounter{table}{-1}
\begin{table*}
\caption[]{(Continued).
}
\begin{tabular}{rrrrrrrrr}\hline
Star &$V$ &$K_s$&$V-K$&[12]&  $F12/F2$ &$\sigma(F12/F2)$ & Field Number \\ \hline
6740   & 13.37   &   -   &   -   &   10.26   &   -   &   -  & F2 \\ 
d-213   & 13.37   &   10.36   &   3.01   &   11.25   &   0.019   &   0.010  & F4 \\ 
6737   & 13.43   &   10.95   &   2.48   &   10.91   &   0.045   &   0.020  & F2 \\ 
2731   & 13.43   &   10.67   &   2.76   &   10.99   &   0.033   &   0.015  & F4 \\ 
1406   & 13.45   &   10.46   &   2.99   &   10.80   &   0.031   &   0.013  & F3 \\ 
d-567   & 13.45   &   -   &   -   &   10.67   &   -   &   -  & F2 \\ 
6623   & 13.46   &   -   &   -   &   10.90   &   -   &   -  & F1 \\ 
e-521   & 13.47   &   -   &   -   &   10.19   &   -   &   -  & F2 \\ 
5423   & 13.48   &   -   &   -   &   11.32   &   -   &   -  & F1 \\ 
8614   & 13.48   &   10.64   &   2.84   &   10.95   &   0.033   &   0.015  & F3 \\ 
2528   & 13.50   &   10.60   &   2.90   &   10.87   &   0.033   &   0.014  & F3 \\ 
2737   & 13.51   &   10.64   &   2.87   &   10.89   &   0.035   &   0.015  & F4 \\ 
2737   & -   &   -   &   -   &   10.96   &   0.033   &   -  & F3 \\ 
1408   & 13.53   &   10.67   &   2.86   &   10.49   &   0.051   &   0.019  & F3 \\ 
1320   & 13.54   &   10.61   &   2.93   &   10.81   &   0.036   &   0.015  & F3 \\ 
1522   & 13.56   &   10.71   &   2.85   &   11.12   &   0.029   &   0.014  & F3 \\ 
6509   & 13.57   &   -   &   -   &   10.59   &   -   &   -  & F1 \\ 
5526   & 13.58   &   -   &   -   &   11.28   &   -   &   -  & F1 \\ 
d-207   & 13.59   &   10.76   &   2.83   &   11.04   &   0.033   &   0.016  & F4 \\ 
2531   & 13.59   &   11.31   &   2.28   &   10.96   &   0.059   &   0.027  & F3 \\ 
5527   & 13.60   &   -   &   -   &   10.63   &   -   &   -  & F1 \\ 
1723   & 13.61   &   10.96   &   2.65   &   10.83   &   0.048   &   0.021  & F3 \\ 
1703   & 13.62   &   10.84   &   2.78   &   11.36   &   0.026   &   0.014  & F3 \\ 
1703   & -   &   -   &   -   &   10.88   &   0.041   &   -  & F4 \\ 
6758   & 13.64   &   -   &   -   &   11.25   &   -   &   -  & F2 \\ 
5645   & 13.64   &   10.88   &   2.76   &   11.17   &   0.033   &   0.017  & F1 \\ 
d-490   & 13.67   &   -   &   -   &   10.80   &   -   &   -  & F2 \\ 
6501   & 13.70   &   -   &   -   &   11.08   &   -   &   -  & F1 \\ 
6614   & 13.70   &   -   &   -   &   11.08   &   -   &   -  & F1 \\ 
6628   & 13.71   &   -   &   -   &   11.18   &   -   &   -  & F1 \\ 
2727   & 13.71   &   11.44   &   2.27   &   11.27   &   0.050   &   0.026  & F4 \\ 
6720   & 13.72   &   11.00   &   2.72   &   11.40   &   0.029   &   0.016  & F2 \\ 
2713   & 13.74   &   11.45   &   2.29   &   10.35   &   0.119   &   0.041  & F4 \\ 
d-514   & 13.74   &   10.88   &   2.86   &   11.26   &   0.030   &   0.016  & F2 \\ 
2519   & 13.76   &   10.93   &   2.83   &   10.63   &   0.056   &   0.022  & F3 \\ 
1407   & 13.78   &   11.00   &   2.78   &   11.19   &   0.036   &   0.018  & F3 \\ 
6715   & 13.81   &   11.16   &   2.65   &   11.35   &   0.036   &   0.020  & F2 \\ 
d-210   & 13.81   &   11.47   &   2.34   &   11.89   &   0.029   &   0.020  & F4 \\ 
1716   & 13.82   &   11.23   &   2.59   &   11.94   &   0.022   &   0.016  & F4 \\ 
6212   & 13.83   &   11.48   &   2.35   &   12.17   &   0.022   &   0.017  & F1 \\ 
\hline
\end{tabular}

\end{table*}

\addtocounter{table}{-1}
\begin{table*}
\caption[]{(Continued).
}
\begin{tabular}{rrrrrrrrr}\hline
Star &$V$ &$K_s$&$V-K$&[12]&  $F12/F2$ &$\sigma(F12/F2)$ & Field Number \\ \hline
2722   & 13.84   &   11.16   &   2.68   &   11.10   &   0.046   &   0.022  & F4 \\ 
1504   & 13.84   &   11.08   &   2.76   &   10.94   &   0.049   &   0.022  & F3 \\ 
1607   & 13.85   &   11.04   &   2.81   &   11.40   &   0.030   &   0.017  & F3 \\ 
7734   & 13.87   &   -   &   -   &   11.49   &   -   &   -  & F2 \\ 
6744   & 13.89   &   11.20   &   2.69   &   11.36   &   0.037   &   0.020  & F2 \\ 
1423   & 13.90   &   11.17   &   2.73   &   11.12   &   0.044   &   0.022  & F3 \\ 
d-218   & 13.91   &   11.54   &   2.37   &   11.75   &   0.036   &   0.024  & F4 \\ 
6611   & 13.91   &   -   &   -   &   11.44   &   -   &   -  & F1 \\ 
6518   & 13.93   &   -   &   -   &   11.37   &   -   &   -  & F1 \\ 
6601   & 13.93   &   -   &   -   &   11.58   &   -   &   -  & F1 \\ 
1411   & 13.94   &   11.64   &   2.30   &   10.92   &   0.083   &   0.037  & F3 \\ 
1516   & 13.94   &   11.18   &   2.76   &   11.71   &   0.027   &   0.017  & F3 \\ 
6762   & 13.94   &   -   &   -   &   11.09   &   -   &   -  & F2 \\ 
8502   & 13.94   &   -   &   -   &   12.54   &   -   &   -  & F3 \\ 
e-522   & 13.95   &   -   &   -   &   11.97   &   -   &   -  & F2 \\ 
2650   & 13.97   &   11.23   &   2.74   &   10.53   &   0.081   &   0.030  & F4 \\ 
6648   & 13.98   &   -   &   -   &   12.02   &   -   &   -  & F1 \\ 
1614   & 13.98   &   12.07   &   1.91   &   11.76   &   0.056   &   0.037  & F3 \\ 
5424   & 13.99   &   -   &   -   &   11.94   &   -   &   -  & F1 \\ 
d-597   & 14.00   &   -   &   -   &   10.75   &   -   &   -  & F2 \\ 
5419   & 14.00   &   11.67   &   2.33   &   11.97   &   0.033   &   0.024  & F1 \\ 
8736   & 14.00   &   -   &   -   &   12.42   &   -   &   -  & F5 \\ 
2527   & 14.00   &   11.90   &   2.10   &   11.37   &   0.069   &   0.038  & F3 \\ 
1617   & 14.01   &   10.83   &   3.18   &   11.43   &   0.025   &   0.014  & F4 \\ 
2746   & 14.01   &   11.81   &   2.20   &   10.45   &   0.148   &   0.053  & F3 \\ 
1715   & 14.01   &   11.91   &   2.10   &   11.46   &   0.064   &   0.037  & F4 \\ 
1715   & -   &   -   &   -   &   11.07   &   0.092   &   -  & F3 \\ 
1717   & 14.01   &   11.41   &   2.60   &   11.66   &   0.034   &   0.021  & F4 \\ 
1526   & 14.01   &   11.73   &   2.28   &   11.25   &   0.066   &   0.034  & F3 \\ 
2735   & 14.01   &   11.82   &   2.19   &   11.23   &   0.073   &   0.038  & F4 \\ 
1712   & 14.02   &   -   &   -   &   11.82   &   -   &   -  & F4 \\ 
6402   & 14.03   &   -   &   -   &   12.22   &   -   &   -  & F1 \\ 
2730   & 14.04   &   11.88   &   2.16   &   10.67   &   0.129   &   0.052  & F4 \\ 
d-212   & 14.04   &   11.94   &   2.10   &   11.57   &   0.060   &   0.036  & F4 \\ 
2622   & 14.04   &   11.75   &   2.29   &   11.02   &   0.085   &   0.040  & F4 \\ 
6610   & 14.05   &   -   &   -   &   13.25   &   -   &   -  & F1 \\ 
6633   & 14.05   &   -   &   -   &   11.26   &   -   &   -  & F1 \\ 
5642   & 14.05   &   11.74   &   2.31   &   11.51   &   0.053   &   0.031  & F1 \\ 
5628   & 14.06   &   -   &   -   &   12.56   &   -   &   -  & F1 \\ 
6630   & 14.06   &   -   &   -   &   11.82   &   -   &   -  & F1 \\ 
\hline
\end{tabular}

\end{table*}

\addtocounter{table}{-1}
\begin{table*}
\caption[]{(Continued).
}
\begin{tabular}{rrrrrrrrr}\hline
Star &$V$ &$K_s$&$V-K$&[12]&  $F12/F2$ &$\sigma(F12/F2)$ & Field Number \\ \hline
1702   & 14.06   &   11.85   &   2.21   &   11.80   &   0.045   &   0.030  & F4 \\ 
7717   & 14.06   &   -   &   -   &   11.08   &   -   &   -  & F2 \\ 
6511   & 14.07   &   -   &   -   &   12.45   &   -   &   -  & F1 \\ 
d-469   & 14.07   &   -   &   -   &   8.97   &   -   &   -  & F2 \\ 
1315   & 14.07   &   12.13   &   1.94   &   11.35   &   0.087   &   0.047  & F3 \\ 
6516   & 14.07   &   -   &   -   &   12.29   &   -   &   -  & F1 \\ 
d-563   & 14.07   &   -   &   -   &   11.04   &   -   &   -  & F2 \\ 
6604   & 14.07   &   -   &   -   &   12.52   &   -   &   -  & F1 \\ 
1429   & 14.08   &   11.79   &   2.29   &   12.19   &   0.029   &   0.023  & F3 \\ 
1754   & 14.08   &   11.86   &   2.22   &   11.09   &   0.086   &   0.042  & F3 \\ 
6603   & 14.09   &   -   &   -   &   11.35   &   -   &   -  & F1 \\ 
5644   & 14.09   &   11.78   &   2.31   &   11.72   &   0.046   &   0.030  & F1 \\ 
d-200   & 14.09   &   11.66   &   2.43   &   11.54   &   0.047   &   0.028  & F3 \\ 
d-249   & 14.12   &   11.09   &   3.03   &   11.02   &   0.045   &   0.021  & F4 \\ 
d-489   & 14.12   &   -   &   -   &   10.47   &   -   &   -  & F2 \\ 
7739   & 14.13   &   -   &   -   &   11.14   &   -   &   -  & F2 \\ 
2747   & 14.16   &   11.51   &   2.65   &   11.49   &   0.044   &   0.026  & F4 \\ 
d-632   & 14.18   &   11.86   &   2.32   &   11.15   &   0.083   &   0.041  & F5 \\ 
1750   & 14.20   &   11.61   &   2.59   &   10.66   &   0.104   &   0.041  & F3 \\ 
6301   & 14.24   &   -   &   -   &   12.44   &   -   &   -  & F1 \\ 
1414   & 14.24   &   11.57   &   2.67   &   11.79   &   0.035   &   0.023  & F3 \\ 
6612   & 14.25   &   11.37   &   2.88   &   10.76   &   0.076   &   0.032  & F1 \\ 
2520   & 14.26   &   11.63   &   2.63   &   11.37   &   0.054   &   0.030  & F3 \\ 
6514   & 14.29   &   -   &   -   &   11.66   &   -   &   -  & F1 \\ 
1732   & 14.31   &   -   &   -   &   10.98   &   -   &   -  & F3 \\ 
6526   & 14.32   &   -   &   -   &   11.50   &   -   &   -  & F1 \\ 
6308   & 14.34   &   11.72   &   2.62   &   11.57   &   0.049   &   0.030  & F1 \\ 
2641   & 14.44   &   11.89   &   2.55   &   10.53   &   0.149   &   0.056  & F3 \\ 
8752   & 14.46   &   12.07   &   2.39   &   11.39   &   0.081   &   0.045  & F5 \\ 
5525   & 14.49   &   -   &   -   &   11.66   &   -   &   -  & F1 \\ 
5425   & 14.49   &   -   &   -   &   11.79   &   -   &   -  & F1 \\ 
1635   & 14.50   &   12.00   &   2.50   &   11.53   &   0.065   &   0.039  & F3 \\ 
8612   & 14.50   &   -   &   -   &   11.15   &   -   &   -  & F3 \\ 
6631   & 14.54   &   -   &   -   &   11.93   &   -   &   -  & F1 \\ 
8731   & 14.57   &   11.67   &   2.90   &   11.09   &   0.072   &   0.035  & F5 \\ 
1318   & 14.58   &   12.86   &   1.72   &   10.64   &   0.328   &   0.129  & F3 \\ 
1751   & 14.62   &   12.10   &   2.52   &   10.70   &   0.154   &   0.062  & F3 \\ 
2652   & 14.63   &   12.18   &   2.45   &   11.24   &   0.101   &   0.052  & F3 \\ 
6529   & 14.64   &   -   &   -   &   12.37   &   -   &   -  & F1 \\ 
5421   & 14.78   &   -   &   -   &   11.82   &   -   &   -  & F1 \\ 
6733   & 14.96   &   -   &   -   &   10.83   &   -   &   -  & F2 \\ 
5640   & 15.03   &   12.44   &   2.59   &   11.65   &   0.089   &   0.056  & F1 \\ 
1640   & 15.05   &   12.07   &   2.98   &   10.98   &   0.116   &   0.053  & F3 \\ 
2427   & 15.07   &   12.76   &   2.31   &   11.43   &   0.148   &   0.084  & F3 \\ 
2651   & 15.46   &   -   &   -   &   11.22   &   -   &   -  & F3 \\ 
6302   & 15.61   &   -   &   -   &   12.04   &   -   &   -  & F1 \\ 
6532   & 15.91   &   -   &   -   &   11.05   &   -   &   -  & F1 \\ 
tuc2251   & -   &   10.00   &   -   &   11.69   &   0.009   &   0.006  & F2 \\ 
tuc1039   & -   &   9.90   &   -   &   10.20   &   0.033   &   0.011  & F3 \\ 
tuc1040   & -   &   9.90   &   -   &   10.12   &   0.035   &   0.011  & F4 \\ 
tuc2198   & -   &   12.46   &   -   &   11.12   &   0.146   &   0.072  & F2 \\ 
d-134,R19   & -   &   7.28   &   -   &   7.36   &   0.040   &   0.004  & F4 & non member\\ 
X3   & -   &   13.35   &   -   &   9.10   &   2.126   &   0.416  & F3 & active galaxy\\ 
\hline
\end{tabular}

\end{table*}

The final list of ISOCAM \mum\ sources obtained from the present observations 
(Table~\ref{Tab:sources}) has been
obtained in the following way. A first list of candidate sources is
provided by the 5~$\sigma$ detections as described in
Sect.~\ref{Sect:ISOCAM}. 
Obviously spurious sources in that list, resulting 
from detector memory effects that were not totally eliminated by the
reduction process, have been discarded. A few faint sources observed
only on the ISOCAM images, but neither on the optical 
(Sect.~\ref{Sect:optical}) 
nor on the DENIS $K$ images 
(Sect.~\ref{Sect:DENIS}) have been discarded as well. 
Table~\ref{Tab:sources} lists the infrared sources, along with their optical
identifiers, ordered by decreasing optical brightness. 

The comparison of the optical images with our list of ISOCAM sources
makes it possible to look for infrared sources with no optical
counterparts. As can be seen on Fig.~\ref{Fig:image}, there are none in
FIELD2. The only source with no optical counterpart found in
the present  survey is the active X-ray galaxy X3 (Verbunt \& Hasinger
1998) observed in FIELD3. Its \mum\ flux amounts to
6.5~mJy. 

One of the important  results obtained by the present study is thus {\it the
absence of bright IR sources without optical counterparts, that could
be associated with dust-enshrouded AGB stars}. 
This result may be used to set an upper limit to the lifetime of dust-enshrouded
AGB stars. To do so, the number of horizontal-branch stars in the 
529 arcmin$^2$ area covered by the fields listed in
Table~\ref{Tab:AOT} is first evaluated. For that purpose, 
positions were taken from the catalogue of Tucholke (1992), 
with $V$ and $B$ magnitudes from
Lee (1977) and Chun \& Freeman (1978). Stars on the horizontal branch 
were selected on the basis of their magnitude and color index
satisfying the conditions $0.75 \le B-V < 0.94$ and $13.9 \le V <
14.2$. With 185 stars fulfilling these criteria, the fraction of
obscured AGB stars to horizontal-branch stars turns out  to be less
than 1/185. Adopting $1.23\times 10^8$~y for the lifetime of stars with $(M/{\rm
M}_\odot, Z) = (0.8,0.004)$ on the horizontal branch \citep{Girardi00}, this
fraction turns into an upper bound of $6.6\times10^5$~y for the duration of the
obscured AGB phase. 
Extrapolating 
to the whole cluster the
absence of obscured AGB stars in the surveyed fields does not make this 
limit much more constraining, since the catalogue of Tucholke (1992) contains 
only about twice more stars on the horizontal branch in the whole cluster.
A limit of a few $10^5$~y is not very meaningful,
since the duration of the high mass-loss episode (``superwind'') necessary to
produce dust-enshrouded AGB stars appears to be of the order of a few $10^4$~y,
according to models of mass-losing AGB stars computed by
Schr\"oder et al. (1999).   
Observations of the
structure of the CO shell around the post-AGB star \object{AFGL 618}
(Meixner et al. 1998) and around some OH/IR objects 
(Delfosse et al. 1997) point towards a duration of a few $10^3$~y or
even less for the superwind phase [the OH/IR objects observed by
Delfosse et al. (1997) are, however,
probably rather
massive ($>5$~\Msun?) and thus not representative of the AGB stars
found in globular clusters].

The brevity of the corresponding evolutionary stage may not be, however, 
the only factor accounting for the absence of obscured AGB stars in 47 Tuc,
since despite similarly adverse statistical expectations, Frogel et
al. (1990) and Tanab\'e et al.
(1997, 1999) nevertheless found such stars in three
different globular clusters belonging to the SMC (NGC 419) or to the
LMC (NGC 1783 and NGC 1978). These clusters belong to classes SWB
IV-VI (Searle et al. 1980) and are characterized by turn-off masses
1.5 -- 1.6~\Msun\ (Nishida et al. 2000), 
much larger than the turn-off mass of 47~Tuc ($\sim
0.9$~\Msun), and ages in the range 1.6 -- 2.0~Gyr.
The obscured AGB stars found by Tanab\'e et al. (1997, 1999) have \Mbol\
$= -4.9$ -- $-5.0$, somewhat larger than the bolometric magnitudes
of the variable stars at the AGB tip in 47 Tuc, which have \Mbol\ between $-3.4$
at minimum and $-4.7$ at maximum (Frogel et al. 1981). The
larger initial masses of AGB stars in the clusters where dust-enshrouded stars
have been found is probably the key to the
existence of such stars in the LMC/SMC globular clusters and not in 47~Tuc. 
Habing (1996) has reviewed the evidence indicating that larger initial masses
imply higher mass-loss rates on the AGB, which in 
turn imply thicker dust shells
[see e.g., Eq.(2b) of Le Sidaner \& Le Bertre (1993) or Bedijn
(1987)]. According to the spectral energy distributions computed by
Bedijn (1987; his Fig.~4), dust-enshrouded stars have optical depths at
9.7~$\mu$m 
in excess of $\sim 5$, and require mass loss rates in excess of
about $10^{-5}$~\Msun~y$^{-1}$ (for typical values of the other relevant
parameters), i.e., in the `superwind' regime.
Schr\"oder et al. (1999) have shown that AGB stars with masses lower
 than 1.1~\Msun\ never reach large enough luminosities to drive a dust-induced 
superwind, and thus have no chance to become dust-enshrouded. 
Finally, the luminosity function of obscured AGB stars in the field 
of the LMC rises steeply at \Mbol\ $\sim -4.5$ (van Loon et al. 1999), which is
about where the AGB ends in 47~Tuc (Frogel et al. 1981).
Therefore, all available evidence points towards the low turnoff mass 
of 47 Tuc
as being the main cause for its lack of dust-enshrouded
stars, due to the absence of a superwind phase in low-mass AGB stars. 

\section{The $F12/F2$ index: dust mass loss as a function of luminosity}

\begin{figure*}[!ht]
\resizebox{\hsize}{!}{\includegraphics{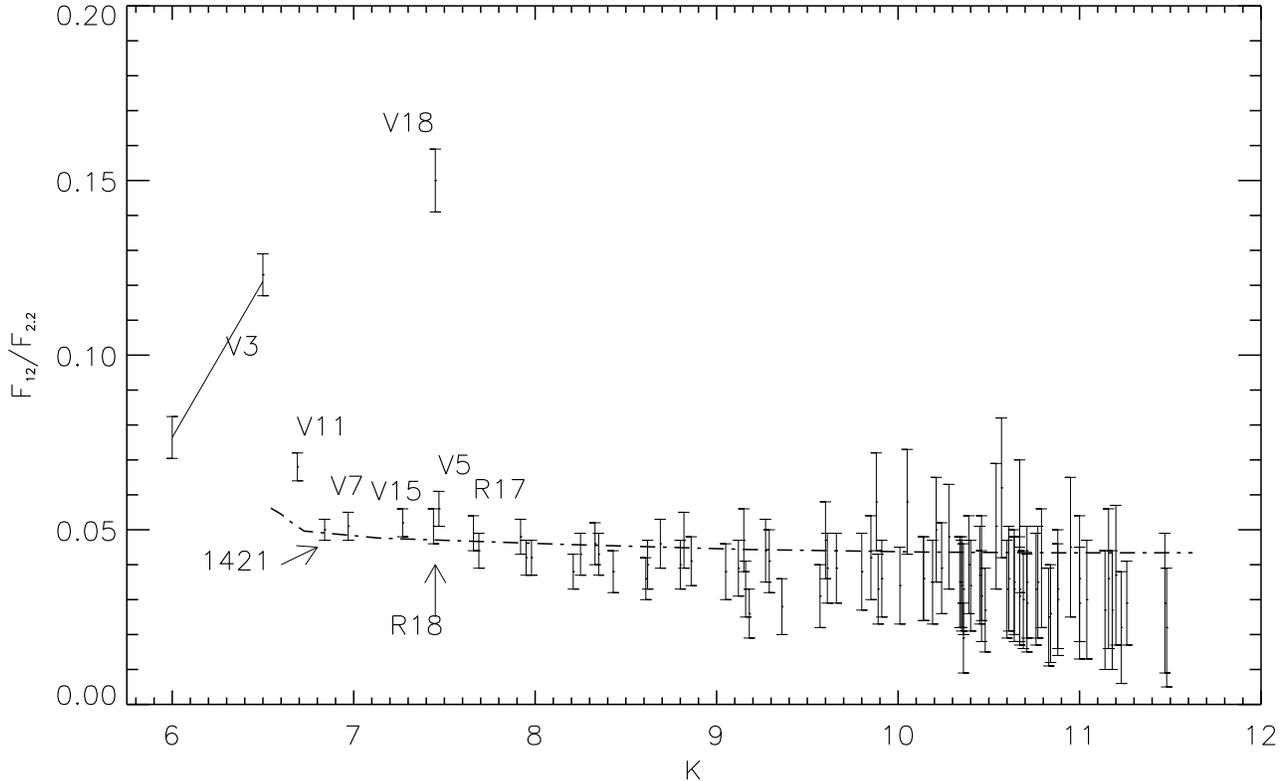}}
\caption[]{\label{Fig:F12F2}
The $F12/F2$ vs $K$ 
diagram for stars detected at 2 and 12 $\mu$m, with
the variable stars identified. The conversion from $K$ magnitudes to fluxes was
performed by adopting a zero-magnitude flux of 665~Jy, following
Fouqu\'e et al. (2000). Note that the distance modulus of 47 Tuc is
13.7 (Gratton et al. 1997). 
The oblique segment for $V3$ translates its variability in the $K$ band only
(see Table~\ref{Tab:IRASISO}), 
as existing data do not allow yet to estimate its variability
range at 12 $\mu$m. 
The dot-dashed line corresponds to the $F12/F2$ ratio
derived from dust-free model atmospheres (see text) 
}
\end{figure*}

As indicated in Sect.~1, the ratio  $F12/F2$ of the 12 to 2 $\mu$m fluxes is 
expected to be a good indicator of dust mass loss.
Fig.~\ref{Fig:F12F2} 
presents the $F12/F2$ vs. $K$ diagram for all the sources detected
at 2 and 12 $\mu$m. As indicated in Sect.~\ref{Sect:ISOCAM}, the error bars
reflect the Poissonian noise on the signal,
 and the uncertainty on the PSF-wing
correction. 
To fix the ideas, the $F12/F2$ ratio expected for dust-free stellar
atmospheres has been estimated from
the stellar atmosphere models of Plez et al. (1992) and Bessell et
al. (1998).
Models of solar metallicity have been used, as only those are
available over the whole temperature range covered by the observed
stars. Nevertheless, a comparison between $F12/F2$ ratios obtained from
(\Teff = 3800~K, $\log g =
-0.5$) models of [Fe/H] = 0 and $-0.6$ (yielding 
$F12/F2 = 0.046$ and 0.045,
respectively)  reveals that metallicity has a negligible impact on the
$F12/F2$ ratio (at least for those model stars with no mass loss).    

The $F12/F2$ ratio for most of the stars is thus consistent with the value
predicted for dust-free photospheres (Fig.~\ref{Fig:F12F2}). 
Only the bright LPV star V3 (the only Mira in Fig.~\ref{Fig:F12F2}, 
with $P = 192$~d)  
and, more surprisingly  the low-amplitude variable V18, do exhibit 
\mum\ excesses. A marginal excess may be present as well for the 
irregular
variable V11. The exact value of the excess is, however, difficult to assess
for V3, as this Mira variable exhibits 0.5-mag variations in the $K$ band
(Frogel et al. 1981), and very likely varies as well in the \mum\
band\footnote{Variations in the 12\,$\mu$m IRAS fluxes by factors as large as
1.6 have been 
reported by Jorissen \& Knapp (1998) for some S-type Mira variables. 
The M-type Mira Z~Cyg also exhibits variations by a factor of 2 at 12\,$\mu$m
as revealed by ISO SWS spectra (Onaka et al. 1999). Finally,  
lightcurves in the $N$ band of O-rich Miras (Le Bertre 1993) reveal variations
by as much as 1~mag (i.e., a factor of 2.5).
These variations may be accounted for by the underlying luminosity and 
temperature variations of the Mira variable.}.
The $F12/F2$ ratio of V3 should therefore ideally be derived from 2 and
12~$\mu$m~fluxes obtained at the same epoch, which is unfortunately not the case
for the value displayed in Fig.~\ref{Fig:F12F2}.
It is very likely, though, that V3 does exhibit at least some excess at \mum.

The 12\,$\mu$m excess observed for V18 is 
more surprising (there is no doubt about its reality as may be assessed from
Fig.~\ref{Fig:image}),
especially when compared to the absence of a similar excess for the stars V7 and
V15, which are located slightly above V18 on the giant branch. 
The origin of the difference between the $F12/F2$ ratios of variable stars of
comparable luminosities must probably be searched for in their variability
properties. 
Both V11 and V18 are noteworthy in that respect. Fox (1982) notes that 
the
irregular variable V11 possibly changed its period (and pulsation mode?) from
about 100~d at the time of Lloyd Evans' observations (Lloyd Evans 1974) to 52~d
about 10~y later. 
Similarly, Fox (1982) suggests that V18 may be of the type of variable
that switches on and off intermittently. In the same vein, Tucholke (1992)
lists two different $B$ magnitudes for V18 ($B = 17.4$ and 13.5 for
entries 2197 and 2198, respectively, which have exactly the same
coordinates); they probably indicate that strong variations occurred
over the time span covered by Tucholke's plates.    
It would be tempting to relate this
behaviour with the sudden ejection of dust shells causing the star to brighten
up at 12~$\mu$m. A similar phenomenon has been proposed for semi-regular
variable stars by Ivezi\'c \& Knapp (1999).   
Note that the large $F12/F2$ ratio of V18 cannot be ascribed to
variations in the $K$ band, since the $K_s$ magnitude estimated from
the raw DENIS frame\footnote{Aperture photometry has been performed on 
  the DENIS frame, and the zero-point has been estimated from a linear 
  regression between the raw DENIS magnitudes and the $K$ magnitudes
  from Frogel et al. (1981) for the stars in common between the two
  samples. The slope of the regression
  line amounts to 1.013 and the maximum relative deviation is 2\%.}
amounts to 7.33, as compared to 7.41 from the {\it Two Micron All Sky 
Survey}
(Skrutskie 1997) and 7.44 obtained almost 20~y earlier by Frogel et al. 
(1981). 
Despite the fact that V28 is also suspected by Fox (1982) of being an
intermittent
variable, this variable star does {\it not}, however,  
show large excesses at \mum\ 
($F12/F2 = 0.067$  with $K = 6.99$ from Fox 1982). 

Nevertheless, Fig.~\ref{Fig:F12F2} shows that the dust mass loss does not
increase
gradually along the AGB, but that it appears suddenly as the AGB star becomes
strongly variable. 
The mass loss rates of the dusty AGB stars in 47~Tuc is not expected
to be quite large, though.
A blind application of the empirical 
relationship between the mass loss rate and the
$K - [12]$ color\footnote{$K - [12] = 
-2.5\;\log[(F_2/F_{12}) (F_{12}(0)/F_{2}(0))]$. In the
definition of the color index $K - [12]$, $F_{2}(0) =
665$~Jy and $F_{12}(0) = 28.3$~Jy are normalization factors ensuring that
stars with no dust excess have $K - [12]$ indices close to 0 (Fouqu\'e 
et al. 2000; IRAS Explanatory Suppl., 1988)}  
(Le Bertre \& Winters 1998) yields 
a total mass loss rate lower than $10^{-8}$~\Msun~y$^{-1}$ for V18!
In fact, this relationship applies to solar-metallicity stars and to the
regime of dust-driven winds, 
when the dust is abundant enough to couple to the gas. Neither
condition is satisfied for the LPVs in 47 Tuc. 
Netzer \& Elitzur (1993) have shown that the coupling between gas and
dust is not achieved below mass loss rates of a few $10^{-8}$~\Msun~y$^{-1}$
(see also Fig.~20 of Jorissen \& Knapp 1998).
Such a low value for the mass loss rate of V18 is also confirmed by
the application of a simple dust radiative-transfer model
encapsulated in Eq.~(6) of Frogel \& Elias (1988, as corrected by
Eq.~(17) of Jura \& Kleinman 1992).
Adopting for V18 $F_{12}({\rm shell}) = 0.07$~Jy (see Table~\ref{Tab:IRASISO},
after subtraction of the
photospheric flux), $v = 4$~km~s$^{-1}$ for the dust outflow
velocity [according to Fig.B1 of van Loon (2000) for variable stars with short
periods], $T_{\rm eff} = 3680$~K and $M_{\rm bol} = -2.96$ (Frogel et
al. 1981) yielding $R = 85$~R$_\odot$, and a grain cross section per unit mass 
$\kappa = 2.3\;10^3$~cm$^2$~g$^{-1}$ at \mum\ for
astronomical silicates   with a density of 3.3 g~cm$^{-3}$ 
(Draine \& Lee 1984)
yields $\dot{M} = \psi\; 2\;10^{-11}$~M$_\odot$~y$^{-1}$, where
$\psi$ is the gas-to-dust ratio. With $\psi$ of the order of
$10^3$, as expected for envelopes with metallicities of 1/5 the solar value 
(solar-metallicity oxygen-rich envelopes typically have $\psi \sim 200$;
e.g., van der Veen 1989, Justtanont et al. 1994), a mass loss rate 
of the order of $10^{-8}$~M$_\odot$~y$^{-1}$ is indeed obtained.
Incidentally, the
$K - [12]$ index of 1.37 for V18 appears quite small in comparison with mass-
losing AGB stars in the solar neighbourhood, where $K - [12]$ falls in
the range 2 -- 6 (corresponding to $0.27 \le F12/F2 \le 11$). 
This result is consistent with the conclusion of 
Sect.~\ref{Sect:list} that
the low-mass
AGB stars of 47~Tuc never reached luminosities high enough to trigger 
dust-driven winds. 

In any case,  the fact that \mum-excesses are observed only for the LPV V3,
for the intermittent variable V18 and possibly for the period-changing variable
V11, is an indication that stellar pulsations
play a key role in triggering dust mass loss. Frogel \& Elias (1988)
reached the same conclusion, based on the presence of 10~$\mu$m
excesses only in globular-cluster LPV stars and not in non-LPV
cluster giants.  
The mass loss-pulsation connection is actually already
well-documented, 
both on observational and theoretical grounds. 
A correlation between the mass loss rate and the pulsation period or
amplitude of variation in the $K$ band has been noticed by various
authors (Whitelock et al. 1987, 1994 for O-rich Miras, 
Claussen et al. 1987 and Groenewegen et al. 1998 for C-rich Miras,
Vassiliadis \& Wood 1993 for a mixture of both). The various model
calculations of mass loss in AGB stars have also shown the key
importance of pulsations for triggering  the mass loss process (see
the discussion in Sect.~\ref{Sect:intro}).

\section{Conclusion}

The main results of the present study are (i)  the absence of dust-enshrouded 
AGB stars in the surveyed fields, (ii) the existence of excess
emission at  \mum\ due to circumstellar dust only for large-amplitude
(or intermittent) variable stars.

The intermittent variable V18 appears to have an excess at 12~$\mu$m
much larger than stars of similar luminosities on the giant branch. A
long-term monitoring of this star spanning several years and covering the
optical and infrared bands would be of great interest to identify the origin
of its anomalous properties.   

\acknowledgements
M. Groenewegen is thanked for his help in the early phase of this project.
D.~Elbaz and H.~Aussel provided invaluable help with the reduction of the
data. The ISOCAM data presented in this paper was analysed using {\it  CIA},
a joint development by the ESA Astrophysics Division and the ISOCAM
Consortium. The ISOCAM Consortium is led by the ISOCAM PI, C. Cesarsky.

\end{document}